\documentstyle[onecolumn,epsf]{mn}

\newif\ifAMStwofonts

\newcommand{\gsim}{\mbox{\raisebox{-1.0ex}{$~\stackrel{\textstyle >}
{\textstyle \sim}~$ }}}
\newcommand{\lsim}{\mbox{\raisebox{-1.0ex}{$~\stackrel{\textstyle <}
{\textstyle \sim}~$ }}}
\newcommand{\psim}{\mbox{\raisebox{-1.0ex}{$~\stackrel{\textstyle \propto}
{\textstyle \sim}~$ }}}

\newcommand{\lmk}{\left(}
\newcommand{\rmk}{\right)}
\newcommand{\lnk}{\left\{ }
\newcommand{\nn}{\nonumber}
\newcommand{\rnk}{\right\} }
\newcommand{\lkk}{\left[}
\newcommand{\rkk}{\right]}
\newcommand{\lla}{\left\langle}
\newcommand{\p}{\partial}
\newcommand{\rra}{\right\rangle}
\newcommand{\beq}{\begin{equation}}
\newcommand{\eeq}{\end{equation}}
\newcommand{\beqa}{\begin{eqnarray}}
\newcommand{\eeqa}{\end{eqnarray}}

\newcommand{\lab}{\label}

\title{Long Term Operation of LISA and
Galactic Close  White Dwarf Binaries} 
\author[Naoki Seto]
       {Naoki Seto \\
 Department of Earth and Space Science,
Osaka University,
Toyonaka 560-0043, Japan\\
seto@vega.ess.sci.osaka-u.ac.jp
}

\begin{document}

\maketitle

\begin{abstract}
The binary confusion noise spectrum at LISA band
 depends strongly on  observational
 period and abundance of  Galactic  close white
 dwarf binaries (CWDBs). 
We have investigated how the number of the  resolved  Galactic
 CWDBs varies with
 operation period of LISA, and  found that the resolved 
number would typically grows by a factor of 5
 when the operation period increases from 1yr to 10yr. We have also made 
 a similar estimation  for 
 number  of CWDBs whose chirp signal can be measured in matched
 filtering analysis. 

\end{abstract}


\section{INTRODUCTION}
If things go well, the Laser Interferometer Space Antenna
(LISA\footnote{http://lisa.jpl.nasa.gov}) will be launched within ten
years.  LISA and ground-based detectors (such as, TAMA300, GEO600, LIGO
and VIRGO) will bring us
fruitful information of our 
universe and era of gravitational-wave astronomy will really start.
 For example, using LISA, we   might detect 
merging 
massive black holes (MBHs) with significant signal to noise ratio (SNR),
 or in-spiraling compact stars (such as, white dwarfs,
neutron stars, 
stellar mass black holes) around  MBHs.  These are very exciting
phenomena  and we might    confirm existence of
black holes, make stringent tests of general relativity and measure
various  interesting parameters of MBHs (Bender et al. 1998).
However, event rate of MBH-MBH binaries is highly unknown ({\it e.g.}
Haehnelt 1994) and
gravitational waves from in-spiraling compact stars  around  MBHs 
might be too much complicated to be detected  by  usual matched filtering
technique (Bender et al. 1998 and
references therein). 

Galactic binaries, such as, neutron star binaries,
cataclysmic binaries, close white dwarf binaries (CWDBs), are guaranteed
sources for LISA (Mironowskii 1965, 
Evans, Iben \& Smarr 1987, Hils, Bender \& Webbink 1990). The
 Galactic CWDBs are  expected to be the dominant  one in the frequency
region from $\sim 10^{-3}$Hz up to several $10^{-2}$Hz (Bender et
al. 1998).  At present  only
less than ten CWDBs have been optically detected, but LISA would find
thousands of CWDBs  with   one year integration (Hils \& Bender 1997,
Bender et al. 1998). Observational analysis
for  abundance and spatial
distribution of Galactic CWDBs would bring us important clues to
understand  formation of 
binary stars and structure of our galaxy ({\it e.g.} Ioka, Tanaka \&
Nakamura 2000, 
Hiscock et al. 2000, see also Oppenheimer et al.
2001). It should be  also noted  
that CWDBs are regarded as one of the likely progenitors of type I
supernova (Iben \& Tutukov 1984, Branch et al. 1995). 
Thus gravitational waves from 
 Galactic CWDBs, one of the guaranteed sources of LISA,   have
rich scientific contents.

An interesting feature of Galactic CWDBs is that they would also 
become an serious  background noise (confusion noise) for measurement
of gravitational wave (Evans et al. 1987). Properties of its 
spectrum as well as the 
estimation errors for  parameters of resolved binaries
  depend strongly on 
observational period.  But  operation period of LISA is
not determined definitely  at present. The {\it Pre-Phase A Report}
(Bender et al. 1998) of
LISA states  
``This (a 1-year-long observing period) is reasonable length of time,
but not the maximum: the nominal mission lifetime is 2yr, but in
principle it might last as long as a decade.''.
Thus it seems interesting to investigate ahead of time 
 how  scientific impacts on CWDBs would 
 change with  operation period of LISA. We mainly investigate the
number of the resolved CWDBs for various observation periods.  
This quantity would be most fundamental for observational analyses
described in the last paragraph. 

This article is organized as follows. In \S 2 we  mention
gravitational wave from CWDBs and a model of their spatial distribution
in our galaxy. In \S 3 we determine the confusion noise curve as a
function of the
observational period and the
abundance of Galactic CWDBs. We also discuss  
signal to noise ratio and the
parameter estimation errors for binaries
 in  matched filtering analysis.
Then in \S 4  we evaluate number of resolved CWDBs for various set of
parameters. \S 5 is devoted to a brief summary.   

\section{Gravitational Wave from Galactic CWDBs}

\subsection{Chirp Gravitational Wave}

Let us consider a binary (masses $M_1$ and $M_2$) with a circular orbit. 
The characteristic amplitude $h_A$ of gravitational wave at a frequency
$f$ is evaluated by  Newtonian quadruple formula as (Thorne 1987)
\beqa
h_A&=&8 \lmk \frac2{15} \rmk^{1/2}\frac{G^{5/3}}{rc^4} (\pi M_c
f)^{2/3}\\ 
   &=&1.2\times 10^{-21}\lmk \frac{M_c}{0.3 M_\odot}  \rmk^{5/3}
    \lmk \frac{100{\rm pc}}{r}  \rmk  \lmk \frac{f}{10^{-3}{\rm Hz}}
\rmk^{2/3}, \lab{an}
\eeqa
where $M_c\equiv (M_1 M_2)^{3/5} (M_1+M_2)^{-1/5}$ is the chirp mass of
the system  and $r$ is the distance to it.  In the above equation we have
taken angular 
average of wave amplitude with respect to  orientation of  sources (Thorne
1987). 
The time profile of a nearly monochromatic  wave is expressed as follows 
\beq
h(t)=h_A \cos\lkk 2\pi\lmk ft+\frac{\dot f}2t^2\rmk+\phi_0 \rkk.\lab{ht}
\eeq
Here $\phi_0$ is an integral constant.  For a purely monochromatic wave we
have  
${\dot f}=0$, but energy loss due to gravitational radiation  or other
effects, such as,  
accelerating motion of the binary or mass exchange between binary stars,
  make modulation of the wave  frequency ${\dot
f}\ne 0$.
We denote ${\dot
f}\ne 0$ as follows
\beq
{\dot f}=\lmk {\dot f} \rmk_{GW}
+\lmk {\dot f} \rmk_{other},
\eeq
where the first term of the right-hand-side represents effect due to
gravitational radiation. 
When the wave frequency $f$ changes significantly within an observational
period $T$, namely
\beq
{\dot f}T/f\gsim O(1) ,
\eeq
we might separate these effects in matched filtering analysis using
their frequency dependence.   But this is not the case for
our analysis for 
 low-frequency waves. Thus we  try to measure $\dot f$ as
the sole  
parameter for the
frequency modulation.  Effects due to motion of LISA would be commented
in \S 3.2.

Our main target is Galactic CWDBs. 
A CWDB is expected to have  an almost circular orbit as a result of
spiral-in process during its  common envelope phase, in contrast to
a double neutron star binary (Ignatiev et
al. 2001).
Before the less
massive star fills its Roche lobe,  gravitational  radiation reaction  is
the dominant process for evolution of orbital parameters of a
binary  (see {\it e.g.} Webbink 1984). 
The energy loss due to gravitational radiation is   given by   
Newtonian quadruple formula and changes the wave frequency as follows
(Thorne 1987)
\beqa
\lmk {\dot f} \rmk_{GW}&=&\frac{96 \pi^{8/3}}5
\frac{G^{5/3}}{c^5} f^{11/3}M_c^{5/3}\\
&=&7.9\times 10^{-19}\lmk \frac{f}{10^{-3}{\rm Hz}} \rmk^{11/3}\lmk
\frac{M_c}{0.3 M_\odot} 
\rmk^{5/3}   {\rm sec}^{-2}  .\lab{chirp}
\eeqa
CWDBs are mainly 
divided into three segments (Webbink 1984, Iben \& Tutukov
1984, Tutukov \& Yungelson 1986). They are He-He systems with total mass 
$0.5-0.75M_\odot$, He-CO systems with total mass 
$0.75-1.45M_\odot$, and CO-CO systems  with total mass 
$1.45-2.4M_\odot$.  Webbink (1984) discussed that  these three
systems have similar abundance (see also Branch et al. 1995).  In this
article  we do not distinguish 
them but  study a single component with chirp mass $M_c=0.3
M_\odot$. More detailed treatment would not significantly 
change our main conclusions.

\subsection{ Number and Spatial Distributions of Galactic CWDBs}
The abundance of Galactic CWDBs has not been
observationally clarified (Hils et al. 1990, Marsh 1995, Marsh,
Dhillon, \& Duck 1995, Bender \& Hils 1997, Knox, Hawkins, \& Hambly
1999). 
  In the frequency region 
where  they evolve by gravitational radiation reaction, the
number of CWDBs per unit frequency $dN/df$ becomes 
\beq
\frac{dN}{df} \propto \lmk \frac{df}{dt}  \rmk^{-1}\propto 
  f^{-11/3}, \lab{k1}
\eeq
from  equation (\ref{chirp}). The above distribution can be also
expressed in an integral form as $N(>f)\propto f^{-8/3}$.

We use  estimation for the abundance of  Galactic CWDBs
 given in Hils \& Bender
(2000) as a reference. 
This estimation is 10\% of   the theoretical estimation by Webbink
(1984).    Hils \& Bender
(2000) commented that the error in this estimation is believe to
 be no more than a factor of ten in either  direction.  
The number distribution  $dN/df$ corresponding to this 10\% estimation
is a factor 10 lower than that given in their figure 4 which represented
the full 100\% estimate based on Webbink (1984). Thus we have an
approximation for this 10\% estimation as 
\beq
\frac{dN}{df}=3.2\times 10^7 \lmk\frac{f }{10^{-2.82}{\rm Hz}}\rmk^{-11/3}
 {\rm
Hz}^{-1} , \lab{k2}
\eeq
at the frequency region $f\ga 10^{-3}$Hz in interest.
In this article   we use the following  non-dimensional quantity
 $N_{282}$ to
characterize the abundance of Galactic CWDBs 
\beq 
N_{282}\equiv \frac1{(3.2\times 10^7 {\rm Hz}^{-1})
}\frac{dN}{df}{\bigg|}_{f=10^{-2.82}{\rm Hz}} ,\nn\lab{n282}
\eeq
or we have
\beq
\frac{dN}{df}=3.2\times 10^7 N_{282} \lmk\frac{f }{10^{-2.82}{\rm
Hz}}\rmk^{-11/3} 
 {\rm
Hz}^{-1} . \lab{k3}
\eeq
For example  we have $N_{282}=1$ for   distribution (\ref{k2}).
Webbink \& Han (1998) discussed CWDBs with various 
models of binary evolution. Their ``standard model'' corresponds to
$N_{282}\sim 2.0$. 
Nelemans, Yungelson, \& Portegies 
Zwart (2001) obtained a similar results.  
Note that $(3.2\times
10^7)^{-1}$Hz is the frequency bin for 1yr integration.  Thus $N_{282}$ 
represents the  mean number of CWDBs within ${\rm 1 yr}^{-1}$ bin at
$f=10^{-2.82}$Hz.

Abundance of extra-Galactic CWDB also  becomes  important to discuss
the binary 
confusion noise in the next subsection. We scale their 
abundance by the same parameter $N_{282}$ defined
 for the Galactic ones (see {\it
e.g.} Bender \& Hils 1997 for discussion). 



Next we briefly discuss the 
 spatial distribution of  Galactic CWDBs.   We
use the standard exponential disk model 
\beq
\rho(R,z)=\rho_0 \exp\lmk -\frac{R}{R_0}\rmk \exp\lmk
-\frac{|z|}{z_0}\rmk, 
\eeq
where $(R,z)$ is the Galactic cylindrical coordinate.  We fix the radial
scale 
length $R_0=3.5$kpc and the disk scale height $z_0=90$pc (Hils, Bender
\& Webbink  1990),  and  assume that the solar system exists at the position
$R=8.5$kpc and 
$|z|=30$pc. As 
the amplitude  
of the gravitational wave is inversely proportional to the distance $r$,
distribution of the CWDBs' distances  $r$ from the solar system becomes
important. With our model parameters, $\sim90\%$ of the
 Galactic CWDBs
are within 
$r\le 
18$kpc,  and $\sim 10\%$ of them are within $r\le 5.0$kpc.
The number of binaries within distance $r$ changes from $\psim r^3$
($r$: smaller than disk thickness $\sim100$pc) to $\psim r^2$ ($r$:
smaller than size of Galaxy  $\sim10$kpc) and $\psim r^{0}$ ($r$:
larger than $\sim 10$kpc). 

\section{Matched Filtering Analysis}

\subsection{Confusion Noise}

The signal of a  detector 
$s(t)=h(t)+n(t)$ 
contains both  the true gravitational wave signal $h(t)$ and the noise
$n(t)$.  
We assume the stationary  noise and define  its power spectrum
$S_n(f)$ by
\beq
\lla \tilde{n}(f)\tilde{n}(f')\rra=\frac12 \delta(f-f')S_n(f),
\eeq
where $\tilde{n}(f)=\int^{\infty}_{-\infty}e^{2\pi i f t}n(t)dt$ is the
Fourier transformation of the noise ${n} 
(t)$. This definition of  $S_n(f)$ corresponds to the
one-sided spectral density (e.g. appendix A of Cutler and Flanagan 1994). 
For notational simplicities we have represented the 
frequency space in the continuum limit (see {\it e.g.} Schutz 1997).

At the LISA band the noise spectrum $S_n(f)$ is constituted by two
terms;  (i) the 
instrumental (detector's) noise $S_{ins}(f)$ and (ii) the confusion noise
$S_{con}(f)$ that is effectively 
caused by  unresolved  sources 
\beq
S_n(f)=S_{ins}(f)+S_{con}(f).
\eeq
The  frequency bin $\delta f$ for an observational period $T$ is
simply given as 
\beq
\delta f=T^{-1}=3.1\times 10^{-8} \lmk \frac{T}{\rm 1 yr}\rmk^{-1} {\rm
Hz}. \lab{bin}
\eeq
The number $dN/df$ of Galactic CWDBs per unit frequency increases at
lower frequency as shown in 
equation (\ref{k1}).
Roughly speaking, when the frequency bin (\ref{bin}) is occupies by more
than one Galactic CWDBs,
   the Galactic confusion noise  $S_{con,G}(f)$ becomes
important ({\it e.g.} Bender \& Hils 1997).  At higher frequencies there
is no more than 
 one Galactic binary and
the confusion noise $S_{con,eG}(f)$ is determined by the extra-Galactic
binaries. 
The shape of the confusion
noises  (including the position of the transition frequency $f_{t}$ where
the Galactic 
noise becomes important)  is determined by the
observational period $T$ and the number  of  binaries. We  discuss them
with  a   method similar to Hiscock   et
al. (2000) who  studied dependence of the noise spectrum
on the abundance and spatial
distribution of CWDBs (see also Phinney 2001). 

Bender \& Hils (2000 and references therein) discussed that the 
confusion 
noise  in the region $10^{-4}<f<10^{-1.5}$Hz is dominated by
 CWDBs.  They
found that in some scenarios (Tutukov \& Yungelson 1996, see also Iben \&
Tutukov 1991) the abundance
$dN/df$ of the Helium Cataclysmics (HeCVs) exceeds that of CWDBs, but
amplitude of confusion noise is expected to be dominated by CWDBs
because of their larger chirp masses than that of HeCVs.  Thus in our
present analysis we 
basically do not consider the effects of HeCVs and study the spectrum
$S_{con} (f)$ as a function of $T$ and abundance of CWDBs (characterized
by 
 $N_{282}$).

 The extra-Galactic noise $S_{con,eG}(f)df$ has the following relation
\beq
S_{con,eG}(f)df \propto ({\rm number~of~binaries~in~} f\sim
f+df)\times({\rm amplitude ~of 
~individual 
~source})^2.
\eeq
From equations (\ref{an}) and (\ref{k1}) we have 
\beq
S_{con,eG}(f)\propto N_{282} f^{-7/3}.
\eeq
The Galactic confusion noise $S_{con,G}(f)$ at frequency  smaller than
the transition 
 region  around the frequency 
$f_{t}$ has the same functional shape $S_{con,G}(f)\propto N_{282}
f^{-7/3}$.  The ratio of two amplitudes
$\sqrt{S_{con,G}(f)/S_{con,eG}(f)}$  is estimated to $3\sim 10$ and the
uncertainty is mainly related to that of star formation rate at high
redshift (Kosenko \& Postnov 1998, Schneider et al. 2001).

Next we evaluate the
transition frequency $f_{t}$  as a function of the parameters $N_{282}$
and $T$ 
\beq 
\frac{dN}{df}{\bigg|}_{f_{t}} \delta f= Y,\label{fty}
\eeq
where $Y$ is a constant of order unity.
 From equation (\ref{k3}) we obtain
\beq
f_{t}\propto (N_{282}T^{-1})^{3/11}. \lab{ft}
\eeq

Now we can  make the confusion noise spectra $S_{con}(f,N_{282},T)$ for
various parameters $(N_{282},T)$ using 
a functional form $S_{con}(f,N_{282,0},T_0)$ given for a specific choice
of parameters $N_{282,0}$ and $T_0$.  We connect two (Galactic and
extra-Galactic) power-law functions $\propto N_{282}f^{-7/3}$ around the
characteristic frequency $f_t$ and straightforwardly obtain the
following relation 
\beq
S_{con}(f,N_{282},T)=\lmk\frac{N_{282}}{N_{282,0}}\rmk \lmk\frac{f_{t,0}}{f_t} \rmk^{7/3}
S_{con}\lmk f \lmk\frac{f_{t,0}}{f_t} \rmk ,N_0,T_0  \rmk ,
\eeq
where we have 
\beq
\frac{f_{t,0}}{f_t} =
\lmk\frac{N_{282,0}T_0^{-1}}{N_{282}T^{-1}}
\rmk^{3/11}
\eeq
from equation (\ref{ft})

For the ``base'' function of  the confusion noise 
$S_{con}(f,N_{282,0},T_0)$ we 
use the result  of  Hils \& Bender (2000) given for
parameters $N_{282}=1$ and $T_0=1$yr (see their figure 5).
The instrumental noise $S_n(f)$ is obtained from their same figure.
In figure 1 we  show the noise spectrum for various choice of parameters
 $(N_{282},T)$.

\begin{figure}
 \epsfxsize=8cm
\epsffile{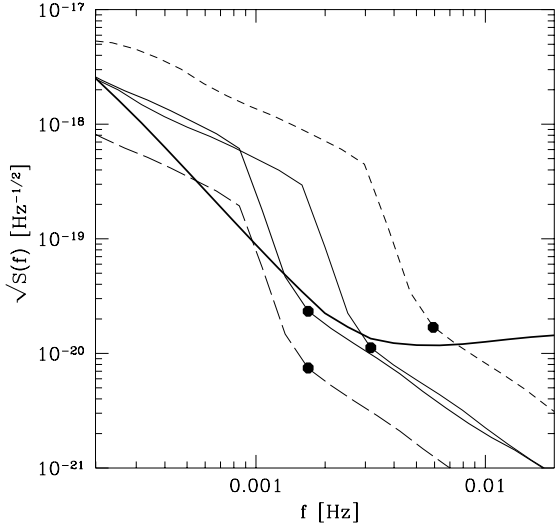}
\caption{ 
The noise spectrum is presented in the form of $\sqrt{S(f)}$.
The instrumental noise of LISA is presented with bold thick line (Hils
 \& Bender 2000).  Other
 lines are confusion noise with various set of parameters $N_{282}$ and
 $T$ (in units of year). The thin solid line corresponds to $(N_{282},
 T)=(1,1)$, thick sold line to (1,10), long-dashed line to (0.1,1) and
 short-dashed line to (10,1).  The filled circles represent the
 frequency $f_t$ defined in eq.[\ref{fb}].}
\end{figure}

Bender \& Hils (1997) gave confusion noise curves 
 $S_{con}(f)$ due to CWDBs for
two different  $N_{282}$.   We find that those  two curves are
reasonably reproduced by each other with the above procedure.  Our
method 
would be quantitatively valid at frequency 
\beq
f> 3.0 \times 10^{-3} \lmk {N_{282}} \rmk^{3/11} \lmk\frac{T}{\rm 1yr}
\rmk^{-3/11}\equiv f_t, \lab{fb}
\eeq
where the confusion
noise is determined by extra-Galactic binaries. We
have put $Y=0.1$ in equation (\ref{fty}) for definition of $f_t$.
The Galactic binaries with $f>f_t$ are not overlapped in the frequency
bin $T^{-1}$ and, in principle, simple to analyzed. In contrast these
with $f<f_t$ can be detected only when they are close to us and have
significant wave amplitude above confusion noise.  Their analysis would
not be straightforward. Here we effectively discuss them  
using the confusion noise
spectrum.

  In
figure 1 
the frequency $f_t$ is shown with filled circles. 
Note that the Galactic 
confusion noise at the transition region for $T=1$yr  vanishes in the
case of 
$T=10$yr. Comparing $T=1$yr and 
$T=10$yr  we find that 
$\sqrt{S_{con}(f)}$ of the latter is smaller about a factor of ten than
the former   at
$f\simeq 
0.0018$Hz in the case of  $N_{282}=1$.
Note that this factor is determined by the quantity
$\sqrt{S_{con,G}(f)/S_{con,eG}(f)}$ as discussed before.

\subsection{SNR and  Estimation Errors  for a  Chirp Signal}
In this subsection we briefly discuss  signal-to-noise ratio (SNR)
and the parameter estimation errors  for
gravitational wave signal in  matched
filtering analysis (Cutler \& Flanagan 1994).
First, we define an inner product of two quasi-periodic waves  $g(t)$ and
$k(t)$  
around a frequency $f_0$ as follows
\beq
(g|k)\equiv \frac{2}{S_n(f_0)}\int_0^T g(t)k(t)dt. \lab{seki}
\eeq
The signal to noise ratio of a gravitational wave $h(t)$ is given
as  
\beq
SNR=(h(t)|h(t))^{1/2}. \lab{snr}
\eeq
When the wave form  $h(t)$ is characterized by some parameters
$\lambda_i$,  magnitude of the 
estimation errors  $\Delta \lambda_i$ is  given by the so called
Fisher information 
matrix $\Gamma_{ij}$  as
\beq
\lla \Delta \lambda_i \Delta\lambda_j \rra=\lmk \frac{\p h}{\p
\lambda_i} \big|  
\frac{\p h}{\p \lambda_j}  \rmk^{-1}=\Gamma_{ij}^{-1} .
\eeq
Position and orientation of  LISA  change in time.
This causes  (i) frequency modulation due to Doppler effect  and  (ii)
variation of
detector's sensitivity due to its rotation (Bender et al. 1998).  In
reality 
we  need to fit 
the direction of a source in the matched filtering analysis (Peterseim
et al. 1997, Cutler 1998), but this is a very troublesome task.  In the
present  analysis we  use  an 
angular averaged sensitivity (effectively a factor of $\sqrt5$
degradation, 
Thorne 1987)
and do not try to fit the direction.
We evaluate  the estimation errors for the
  three parameters of wave form  $\lambda_i=\lnk f,{\dot
f},\phi_0 \rnk$ in equation (\ref{ht}). The  chirp signal $\dot
f$ is very important  from astronomical point of  views
 (see {\it e.g.}
Schutz 1986, 1989).
 This parameter is related to  secular effects of gravitational wave
 during the whole observational
period $T$ 
and distinct from annual effects related to the direction of a source
described above. 
However,  we should notice  that due to correlation in  the 
Fisher information
matrix the actual 
parameter estimation errors
 $\Delta \lambda_i$ would be worse than our results.
Thus we must be cautious to discuss the
 measurement error  of the chirp signal
$\dot f$.

Now we can  calculate  SNR and the estimation errors  $\Delta
\lambda_i$ for  
$\lambda_i=\lnk f,{\dot
f},\phi_0 \rnk$.  Using the time function (\ref{ht}) and equations (\ref{seki}) and
(\ref{snr}) we 
obtain  the following results for a  binary
\beq
SNR=\frac{h_A}{\sqrt{S_n(f)}}T^{1/2},\label{snr2}
\eeq
and the information matrix $\Gamma_{ij}$ gives the root-mean-square
values of the errors as 
\beqa
\Delta {\dot f}&=&\frac{3 \sqrt{5}}{\pi}\frac{T^{-2}}{ SNR}, \lab{delf}\\ 
\Delta {f}&=&\frac{4 \sqrt{3}}{\pi}\frac{T^{-1}}{ SNR},\\
\Delta {\phi_0}&=&\frac3{ SNR}.
\eeqa
For comparison with equation (\ref{chirp}) we can write equation
(\ref{delf}) in the 
following form
\beq
  \Delta {\dot f}   =4.3\times 10^{-19} 
\lmk\frac{SNR}{100}\rmk^{-1}
\lmk\frac{T}{10{\rm yr}}\rmk^{-2}   {\rm sec}^{-2}. \lab{delf2}
\eeq

At a frequency $f$ where  the noise $S_n(f)$  does not depend  on the
observation period $T$  (as in the case of $f\gg f_t$: see figure 1), the
above 
quantities have the following relations   
\beqa
SNR &\propto& T^{1/2},\lab{snr3}\\
\Delta {\dot f}&\propto&
T^{-5/2},\lab{cp}\\ 
\Delta {f}&\propto& T^{-3/2},\\
\Delta {\phi_0}&\propto& T^{-1/2}.
\eeqa
Relation (\ref{snr3})
 is a well known result.  Note that the error (\ref{cp}) for the chirp signal becomes
significantly smaller for  longer  integration time  $T$. Considering the
correlation of the Fisher matrix discussed before, the amplitude of
error  $\Delta 
{\dot f}$ might be larger than our estimation (\ref{delf}). But the asymptotic time
dependence $\Delta {\dot f} \propto T^{-5/2}$ would
 be same for integration time  $T$ much larger than time scale ($1$yr) of 
annual modulation  due to rotation of the detector.   

\section{Resolved CWDBs}
In figure 2  we show the number of the Galactic CWDBs resolved with $SNR>
10$.   We take our model parameters for abundance of Galactic CWDBs at
 $N_{282}=0.1,~1.0$ and 10  and for 
observational period at $T=1$yr and 10yr. 
We present the number of the 
resolved CWDBs within frequency bin $f_0<f<1.26
f_0(=10^{1/10}f_0)$ and  explain  our results mainly
using the figure (the upper right panel of Fig.2) given for $N_{282}=1.0$. 

Firstly, we should notice that all the 
Galactic CWDBs with $f\ga 3\times 10^{-3}$Hz are  resolved with $SNR\ge
10$.  Thus we have $N(f_0<f<1.26 f_0)\propto f_0dN/df|_{f_0}\propto
f_0^{-8/3}$ for 
CWDBs at   higher  frequency. Secondly, according to our analysis based
on the confusion noise spectrum,  the number of resolved CWDBs
increases significantly 
at frequency region dominated by Galactic
confusion noise ($f\lsim 10^{-3}$Hz)
 for $T=10$yr comparing
with $T=1$yr.  In figure 3 we plot the distances  $r$ of CWDBs measured
with $SNR=10$ in the case of $N_{282}=1$. At
$ f<f_t$ the distance $r$ suddenly  becomes small due to the Galactic
confusion 
noise.  As the observational period increases, the frequency $f_t$
decreases and we can observe more CWDBs. 
 In table 1 we show the total 
 number of resolved CWDBs for various parameters. We also count CWDBs
only with $f>f_t$. We can understand that  resolved 
 CWDBs mainly  belong to this
group that do not overlap with other Galactic CWDBs.  The total number
for 
$T=10$yr  is
$4\sim 6$ times larger than for  $T=1$yr. As expected, this
improvement is more significant for larger $N_{282}$.
Note that the peak of the distribution $N(f_0<f<1.2f_0)$ corresponds to
the characteristic frequency $f_t$ for $N_{282}=1$ and 10. This means
that we have $N(>f_t)\psim f_t^{-8/3}\psim T^{8/11}$, and the factor
$\sim 5$  mentioned above 
is roughly given by $10^{8/11}\sim 5$.

A  He+He white dwarf binaries coalesce at $f\sim 0.015$Hz and the
He+CO at $f\sim 0.03$Hz. The coalescence frequency of CO+CO white dwarf 
is close to $0.1$Hz (Bender \& Hils 1997). For simplicities we do not
take into account of these cut-off frequencies.

Cutler (1998) studied LISA's angular resolution $\Delta \Omega$
 for binaries with
monochromatic gravitational waves (see his table 1). The resolution
depends 
on the orientations of binaries  as well as  the directions to them.
Typical estimation errors  with 1yr observation  become $\Delta \Omega\sim
3\times 10^{-2}$[sr] at  $f=10^{-3}$Hz and   $\Delta \Omega\sim
1\times 10^{-3}$[sr] at  $f=10^{-2}$Hz (both for binaries with $SNR=10$).

Next we investigate the number of CWDBs whose chirp signal due to
gravitational radiation reaction (eq.[\ref{chirp}])  can be measured. We
define a 
parameter 
$C$ 
by  the following equation 
\beq
\lmk  {\dot f} \rmk_{GW} = C \lmk  {\Delta\dot f} \rmk. \lab{res}
\eeq 
For reference we take $C>10$ as a detection criteria for the chirp
signal by the gravitational radiation reaction.
 This threshold means that the signal of magnitude 
$(  {\dot f})_{GW}$ (eq.[\ref{chirp}]) is measured within
 $10\%$ accuracy in  matched
filtering analysis.  The number
distribution  $N(f_0<f<1.26 f_0)$ of
CWDBs with $C>10$ is shown in figure 2.
Note that CWDBs with 
$C>10$ has $SNR>10$ 
 for reasonable choice of $T$. We can directly
confirm this using equations 
 (\ref{chirp})  and (\ref{delf2}). 
In table 1 we show the total number $N_{chirp}$ with $C>10$. 
The number of CWDBs with $C>10$ increase significantly for long term 
integration. 
The  total
number of CWDBs for $T=10$yr  is  $\sim40$ times larger than $T=1$yr.

As explained earlier, the actual estimation error   $\Delta {\dot
f}$ might be larger than  our evaluation (\ref{delf}) due to correlation with
other errors, such as, angular position of the source. Here we discuss
how the  total number $N_{chirp}$  depends on the observational period
$T$  assuming the following relation from equation (\ref{delf})
\beq
\Delta {\dot f}=A\times (SNR)^{-1} T^{-2},
\eeq
with a normalization constant $A$ that would be
larger than $3\sqrt5/\pi$ in equation
(\ref{delf}). This 
parameter $A$ is effectively 
absorbed to the threshold value $C$.  We can expect  following two
points  from figure 2 (i) the
total number $N_{chirp}$ would be mainly determined by $f_{max}$ where
the distribution $fdN/df$ takes maximum value, and (ii) it  also departs
from the 
asymptotic behavior $\propto N_{282}f^{-8/3}$ at high frequency region.
The former point (i) represents  that number $N_{chirp}$ of resolved binary becomes
\beq
N_{chirp}\psim N_{282} f_{max}^{-8/3}.\lab{nfmax}
\eeq
As the effects of the confusion noise is very small around $f\simeq
f_{max}$,  the total number 
$N_{chirp}$ simply scales as $N_{chirp}\psim N_{282}$ in contrast to
the previous case of the $SNR$ threshold.
  The
latter point (ii)
means that the distance $r$ of CWDBs for a given  threshold $C$
corresponds to  
$r\simeq$(size of Galaxy) at $f=f_{max}$.
Then we can relate $T$ and $f_{max}$ using equations  (\ref{delf}) and 
(\ref{res}) as follows
\beq
\frac{T^{-5/2}}{f_{max}^{2/3}} \sqrt{S_n(f_{max})}\propto f_{max}^{11/3},
\eeq
and obtain 
\beq
f_{max} \simeq f_{max} S_n(f_{max})^{-3/26} \propto T^{-15/26},
\eeq
where we have used $S_n(f)\simeq S_{ins}(f)\psim f^0$ in the frequency
region relevant for the present analysis (see figure 1).
 Using the
relation (\ref{nfmax}) we 
finally obtain 
\beq
N_{chirp}\propto  N_{282}  T^{20/13}.
\eeq
Thus a long observational period $T$ is crucial to increase the total
number $N_{chirp}$.

\begin{figure}
 \epsfxsize=10cm
\epsffile{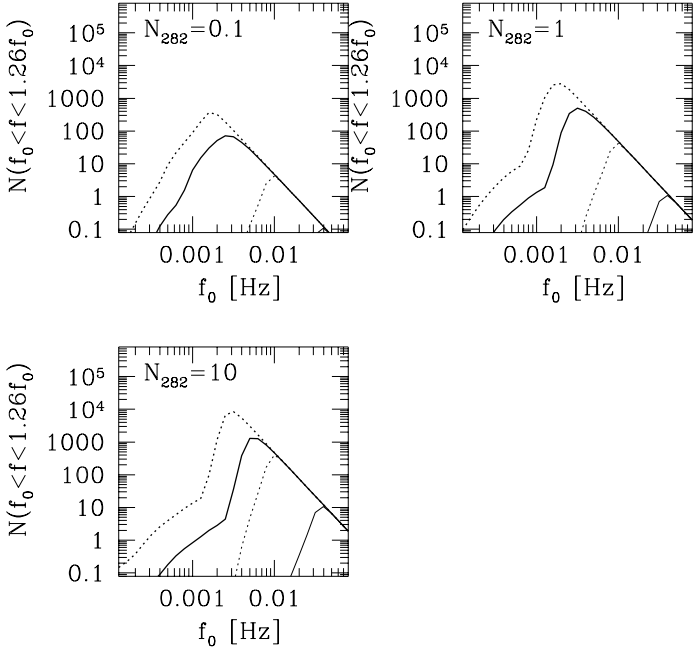}
\caption{Number distribution of resolved Galactic CWDBs within
 frequency bin $f_0<f<1.26f_0$ for parameters $N_{282}=0.1,1$ and 10.
Solid lines represent results for $T=1$yr and dotted lines for $T=10$yr.
 Thick lines are number of resolved CWDBs with $SNR>10$, and thin lines
 with $R>10$ (chirp signal). }
\end{figure}

\begin{figure}
 \epsfxsize=8cm
\epsffile{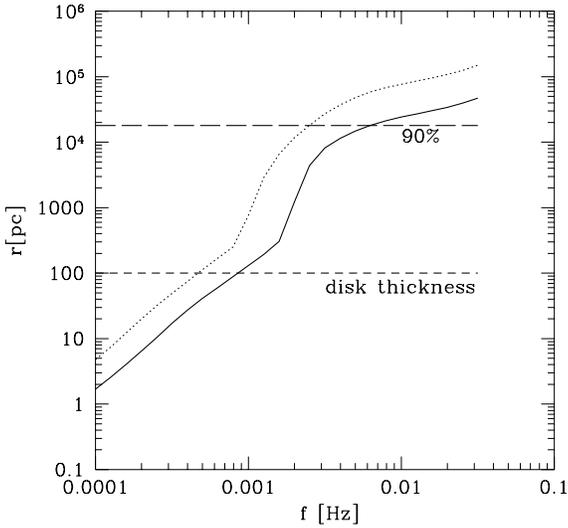}
\caption{Distance $r$ of the Galactic 
 CWDBs resolved with $SNR>10$. We fix the
 abundance 
by $N_{282}=1$.  The sold line represents results for $T=1$yr and the
 dotted line for $T=10$yr.}
\end{figure}

\begin{table}
\caption{number of the resolved Galactic CWDBs}

\begin{tabular}{ccccccc}
 ($N_{282}, T$[yr])  &(0.1,1)& (0.1,10) & (1,1)& (1,10)& (10,1)&(10,10) \\
 $SNR>10$  ~~ &~~ 350~~ &~~  1460  ~~ &~~1950 ~~ &~~10700~~ &~~4810
 ~~ &~~28500 \\
 $SNR>10$, $f>f_t$  ~~ &~~ 325~~ &~~  1420  ~~ &~~1610 ~~ &~~9340~~ &~~3810 ~~ &~~22500 \\
 $C>10$  ~~ &~~ 0.34~~ &~~ 13~~ &~~3.4  ~~ &~~130~~&~~34  ~~ &~~1200\\
\end{tabular}

\medskip
$N_{282}$ represents abundance of the Galactic CWDBs as defined in
 equation (\ref{n282}) and $T$ is the operation period of LISA in units
 of year. $C$ is the resolution threshold for chirp signal
 (eq.[\ref{res}]). 
 \end{table}

\section{Summary}
The Galactic close white dwarf binaries (CWDBs) are one of the
guaranteed sources of LISA. Gravitational wave astronomy for the
Galactic CWDBs  would bring us
important  observational facts to understand binary formation, galactic
structure,  progenitor of type I supernova and so on. 
But gravitational waves from the Galactic
 CWDBs would also become a serious noise (called
confusion noise) below the typical  frequency $f_t(\propto T^{-3/11})$
(eq.[\ref{fb}])  where the effective frequency bin $T^{-1}$ is occupied
by more than one Galactic  binaries ($T$: observational period).  
Apparently  the confusion noise spectrum  depends strongly on the
observational period $T$.

We have investigated number of the resolved
Galactic CWDBs  as a function of $T$ using the
exponential disk model. 
We found that the number $N$ of the resolved CWDBs with $SNR>10$ increases
 $\psim T^{8/11}$ as a function of observational period and it becomes
about a factor of 5 larger  for $T=10$yr comparing with $T=1$yr.
We also studied the number of CWDBs whose  chirp signal can be measured
with matched filtering method.
The chirp signal is one of the most fundamental signal for gravitational
wave astronomy.         Using a rough estimation
we have shown that the number would   grow strongly ($\psim T^{20/13}$)
with 
period $T$.  From $1$yr  to $10$yr operation  the number  can
increase a factor of 
40 for a typical model of  Galactic CWDBs. 

\section*{Acknowledgments}
The author would like to  thank the referee P. L. Bender for helpful
comments to improve manuscript and  K. Ioka  for useful discussions.
This work is  supported by Japanese  Grant-in-Aid for Science Research
Fund of the Ministry of Education, Sports,  Culture, Science and
Technology  Grant
No. 0001461.

\end{document}

\begin{figure}
\epsscale{1}
\plotone{curve.eps}
\caption{The noise spectrum is presented in the form of $\sqrt{S(f)}$.
The instrumental noise of LISA is presented with bold thick line (Hils
 \& Bender 2000).  Other
 lines are confusion noise with various set of parameters $N_{282}$ and
 $T$ (in units of year). The thin solid line corresponds to $(N_{282},
 T)=(1,1)$, thick sold line to (1,10), long-dashed line to (0.1,1) and
 short-dashed line to (10,1).  The filled circles represent the
 frequency $f_t$ defined in eq.[\ref{fb}].}
\end{figure}

\begin{figure}
\epsscale{1}
\plotone{total.eps}
\caption{Number distribution of resolved Galactic CWDBs within
 frequency bin $f_0<f<1.26f_0$ for parameters $N_{282}=0.1,1$ and 10.
Solid lines represent results for $T=1$yr and dotted lines for $T=10$yr.
 Thick lines are number of resolved CWDBs with $SNR>10$, and thin lines
 with $R>10$ (chirp signal). }
\end{figure}

\begin{figure}
\epsscale{1}
\plotone{dis.eps}
\caption{Distance $r$ of the Galactic 
 CWDBs resolved with $SNR>10$. We fix the
 abundance 
by $N_{282}=1$.  The sold line represents results for $T=1$yr and the
 dotted line for $T=10$yr.}
\end{figure}


\begin{thebibliography}{99}

\bibitem{lisa}
Bender, P.\ L.  {\it et al.},
{\it LISA Pre-Phase A Report,} Second edition, July 1998 


\bibitem[Bender \& Hils(1997)]{1997CQGra..14.1439B} Bender, P.\ L.\ \& 
Hils, D.\ 1997, Classical Quantum Gravity, 14, 1439 

\bibitem[Branch et al.(1995)]{1995PASP..107.1019B} Branch, D., Livio, M., 
Yungelson, L.\ R., Boffi, F.\ R., \& Baron, E.\ 1995, PASP, 107, 1019 

\bibitem[Cutler \& Flanagan(1994)]{1994PhRvD..49.2658C} Cutler, C.\ \& 
Flanagan, {\' E}. E.\ 1994, Phys.Rev.D, 49, 2658 

\bibitem[Cutler(1998)]{1998PhRvD..57.7089C} Cutler, C.\ 1998, Phys.Rev.D, 57, 
7089 

\bibitem[Evans, Iben, \& Smarr(1987)]{1987ApJ...323..129E} Evans, C.\ R., 
Iben, I.\ J., \& Smarr, L.\ 1987, ApJ, 323, 129 

\bibitem[Haehnelt(1994)]{1994MNRAS.269..199H} Haehnelt, M.\ G.\ 1994, 
MNRAS, 269, 199 

\bibitem[Hils \& Bender(2000)]{2000ApJ...537..334H} Hils, D.\ \& Bender, 
P.\ L.\ 2000, ApJ, 537, 334 

\bibitem[Hils, Bender, \& Webbink(1990)]{1990ApJ...360...75H} Hils, D., 
Bender, P.\ L., \& Webbink, R.\ F.\ 1990, ApJ, 360, 75 

\bibitem[Hiscock, Larson, Routzahn, \& Kulick(2000)]{2000ApJ...540L...5H} 
Hiscock, W.\ A., Larson, S.\ L., Routzahn, J.\ R., \& Kulick, B.\ 2000, 
ApJ, 540, L5 

\bibitem[Iben \& Tutukov(1984)]{1984ApJS...54..335I} Iben, I.\ \& Tutukov, 
A.\ V.\ 1984, ApJS, 54, 335 

\bibitem[Iben \& Tutukov(1991)]{1991ApJ...370..615I} Iben, I.\ J.\ \& 
Tutukov, A.\ V.\ 1991, ApJ, 370, 615 

\bibitem[]{} Ignatiev, V. B. et al.  2001,  preprint  (astro-ph/0106299)

\bibitem[Ioka, Tanaka, \& Nakamura(2000)]{2000ApJ...528...51I} Ioka, K., 
Tanaka, T., \& Nakamura, T.\ 2000, ApJ, 528, 51 

\bibitem[Knox, Hawkins, \& Hambly(1999)]{1999MNRAS.306..736K} Knox, R.\ A., 
Hawkins, M.\ R.\ S., \& Hambly, N.\ C.\ 1999, MNRAS, 306, 736 


\bibitem[Kosenko \& Postnov(1998)]{1998A&A...336..786K} Kosenko, D.\ I.\ \& 
Postnov, K.\ A.\ 1998, A\&AS, 336, 786 

\bibitem[Marsh(1995)]{1995MNRAS.275L...1M} Marsh, T.\ R.\ 1995, MNRAS, 
275, L1 

\bibitem[Marsh, Dhillon, \& Duck(1995)]{1995MNRAS.275..828M} Marsh, T.\ R., 
Dhillon, V.\ S., \& Duck, S.\ R.\ 1995, MNRAS, 275, 828 



\bibitem[Mironovskii(1965)]{1965SvA.....9..752M} Mironovskii, V.\ N.\ 1965, 
Soviet Astronomy, 9, 752 

\bibitem[Nelemans, Yungelson, \& Portegies 
Zwart(2001)]{2001A&A...375..890N} Nelemans, G., Yungelson, L.~R., \& 
Portegies Zwart, S.~F.\ 2001, A\&A, 375, 890 


\bibitem[Oppenheimer et al.(2001)]{2001Sci...292..698O} Oppenheimer, B.\ 
R., Hambly, N.\ C., Digby, A.\ P., Hodgkin, S.\ T., \& Saumon, D.\ 2001, 
Science, 292, 698 


\bibitem[Peterseim, Jennrich, Danzmann, \& 
Schutz(1997)]{1997CQGra..14.1507P} Peterseim, M., Jennrich, O., Danzmann, 
K., \& Schutz, B.\ F.\ 1997, Classical Quantum Gravity, 14, 1507 

\bibitem[]{} Phinney, E. S.  2001,  preprint  (astro-ph/0108028)


\bibitem[Schneider, Ferrari, Matarrese, \& Portegies 
Zwart(2001)]{2001MNRAS.324..797S} Schneider, R., Ferrari, V., Matarrese, 
S., \& Portegies Zwart, S.\ F.\ 2001, MNRAS, 324, 797 


\bibitem{Schutz:1986}
Schutz,  B.\ F.
1986, Nature\ { 323}, 310 

\bibitem[Schutz(1989)]{1989CQGra...6.1761S} Schutz, B.\ F.\ 1989, Classical 
Quantum Gravity, 6, 1761 



\bibitem[Schutz(1997)]{1997fps..conf..265S} Schutz, B.\ 1997, ESA SP-420: Fundamental Physics in Space, 265




\bibitem{thorne}
Thorne, K.\ S.  in
{\it 300 Years of Gravitation,} edited by S. ~W. ~Hawking and W. ~Israel
 (Cambridge,  England, 1987), pp.330-458.

\bibitem[Tutukov \& Yungelson(1986)]{1986SvA....30..598T} Tutukov, A.\
 V.\  
\& Yungelson, L.\ R.\ 1986, Soviet Astronomy, 30, 598 

\bibitem[Tutukov \& Yungelson(1996)]{1996MNRAS.280.1035T} Tutukov, A.\ \& 
Yungelson, L.\ 1996, MNRAS, 280, 1035 

\bibitem[Webbink(1984)]{1984ApJ...277..355W} Webbink, R.\ F.\ 1984, ApJ, 
277, 355 


\bibitem[Webbink \& Han(1998)]{1998lain.conf...61W} Webbink, R.~F.~\&
 Han,  Z.\ 1998, AIP Conf.~Proc.~456: Laser Interferometer Space
 Antenna, Second  
International LISA Symposium on the Detection and Observation of 
Gravitational Waves in Space, 61 










\end{thebibliography}
\end{document}